\documentclass[11pt]{article}

\newtheorem{condition}{\bf Condition}[section]

\usepackage[top=1in, bottom=1in, left=1.25in, right=1.25in]{geometry}
\usepackage{amsmath}
\usepackage{amsbsy}
\usepackage{amscd}
\usepackage{amsopn}
\usepackage{amstext}
\usepackage{amsxtra}
\usepackage{mathrsfs}
\usepackage{graphicx}
\usepackage{epstopdf}
\usepackage{epsfig}
\usepackage{authblk}
\usepackage{cite}
\usepackage{url}
\usepackage{hyperref}
\usepackage{amsmath}
\usepackage{amsbsy}
\usepackage{amscd}
\usepackage{amsopn}
\usepackage{amstext}
\usepackage{amsxtra}
\usepackage{color}
\usepackage{lscape}

\newcommand{\NB}{\mbox{\sf NegB}}

\newcommand{\beginsupplement}{%
        \setcounter{table}{0}
        \renewcommand{\thetable}{S\arabic{table}}%
        \setcounter{figure}{0}
        \renewcommand{\thefigure}{S\arabic{figure}}%
     }

\begin{document}

\title{Calibrating COVID-19 SEIR models with time-varying effective contact rates}

\author{James~P.~Gleeson$^{1,3,4,5}$,  Thomas Brendan Murphy$^{2,3,5}$,  Joseph~D.~O'Brien$^1$, Nial Friel$^{2,3}$, Norma Bargary$^{1,3,4}$, David J.~P.~O'Sullivan$^1$}
\affil{$^1$ MACSI, Department of Mathematics and Statistics, University of Limerick, Ireland.\\
$^2$ School of Mathematics and Statistics, University College Dublin, Ireland.\\
$^3$ Insight Centre for Data Analytics, Ireland\\
$^4$ Confirm Centre for Smart Manufacturing, Ireland\\
$^5$ Irish Epidemiological Modelling Advisory Group (IEMAG).}

\date{4 June 2021}




\maketitle

\begin{abstract}
We describe the population-based SEIR (susceptible, exposed, infected, removed) model developed by the Irish Epidemiological Modelling Advisory Group (IEMAG), which advises the Irish government on COVID-19 responses. The model assumes a time-varying effective contact rate (equivalently, a time-varying reproduction number) to model the effect of non-pharmaceutical interventions. A crucial technical challenge in applying such models is their accurate calibration to observed data, e.g., to the daily number of confirmed new cases, as the past history of the disease strongly affects predictions of future scenarios. We demonstrate an approach based on inversion of the SEIR equations in conjunction with statistical modelling and spline-fitting of the data, to produce a robust methodology for calibration of a wide class of models of this type.
\end{abstract}

\section{Introduction} \label{sec:1}
The Irish Epidemiological Modelling Advisory Group (IEMAG) was established in March 2020 to provide expert advice to Ireland's Chief Medical Officer and National Public Health Emergency Team  on COVID-19 responses. As part of a suite of mathematical and statistical modelling tools, we developed a population-level susceptible-exposed-infected-removed (SEIR) model based on multiple compartments  \cite{Hethcote2000mathematics,vynnycky2010introduction,Kucharski20}. Many groups have used such models to aid in scenario-based planning for pandemic responses, e.g. \cite{prem2020effect,radulescu2020management,he2020seir}. Although population-level SEIR models use a number of simplifying assumptions such as a fully-mixed and homogeneous population---a simplification that is avoided by more complex age-cohorted or agent-based models \cite{Andrade2020,aleta2020modelling,aleta2020data,Colizza,Hunter}, for example---they enable rapid analysis of potential policy interventions and can give clear quantification of the uncertainty due to limited knowledge of virus parameters. Calibration of such models to the observed data---in our case, the number of daily cases of COVID-19 in Ireland---is an important technical challenge that is heightened by the noisy nature of the data and the uncertainty in parameter estimation. In this paper we describe the SEIR model used by IEMAG and give a detailed description of the calibration algorithm, an early version of which appeared in the technical report~\cite{TechNote2}. We emphasise the uncertainty quantification that is enabled by this approach and we highlight the adaptability of the calibration framework, which enables it to be applied---under certain conditions that we examine in detail---to other models that may be required for future pandemics.

The calibration of SEIR models to noisy data has attracted much attention both before and during the COVID-19 pandemic. Our basic assumption is that the effective contact rate $\beta$ of the model can be considered as time-varying to model the impact of non-pharmaceutical interventions such as working from home, closure of schools and universities, lockdown, etc.; the challenge lies in estimating this time-varying $\beta(t)$. Bayesian and inverse-problem methods for parameter estimation are well-established \cite{Chatzilena2019,Gaburro}, but these usually assume that all parameters are constant in time, or allow only piecewise-constant variations in $\beta$, usually at a predefined set of breakpoints (e.g., the dates that movement restrictions are changed). In contrast, we follow the direction of Mummert~\cite{mummert2013studying}, who showed that a time-varying effective contact rate can be found for SIR systems by an exact inversion of the governing differential equations of the model. In extending this concept, we generalise to a range of models and derive conditions on the model structure, and on the smoothness of the data-fitting function, that are required for this approach to be successful. The smoothness conditions are satisfied by statistical models for data fitting, of which we focus here on the negative binomial generalized additive model (GAM). We note that Goswami et al.~\cite{Goswami} have applied similar ideas to invert SEIR models for COVID-19 data, but they do use the raw (unfitted) data and therefore occasionally obtain negative estimates for $\beta(t)$. They can obtain a smoothed (and non-negative) form by a polynomial fit to the recovered $\beta$ values; in contrast, we use a smooth fit to the data and do not require any postprocessing of the $\beta(t)$ values.

The remainder of this paper is structured as follows. In Section~\ref{sec:IEMAGSEIR2} we introduce the SEIR model and discuss methods for its numerical solution. Section~\ref{sec:Calibration} presents our main results on calibration, including the GAM model and the algorithm (and conditions) for determining the time-varying effective contact rate. In Section~\ref{sec:examples} we give examples of the application of the algorithm and model to investigate scenarios and to demonstrate its applicability to more complex models, such as SEIR models that include vaccination. We draw conclusions in Section~\ref{sec:conclusion} and details of calculations and model properties are included in the Supplementary Material file.

\section{The IEMAG SEIR model}  \label{sec:IEMAGSEIR2}
Population-level SEIR models \cite{Hethcote2000mathematics,vynnycky2010introduction} assume fully-mixed, homogeneous populations. Despite this simplification, they provide useful information for scenario-based planning, with the potential for further extension of the structure (e.g., to dis-aggregated age cohorts) in more advanced models.


\subsection{Model structure}
At each moment of time, every individual in the population is considered to be in one of a discrete number of compartments. The structure of the compartments, and the timescales for individuals to move in and out of compartments, are based on the current understanding of the epidemiology of COVID-19, as evidenced by the extensive literature review and evidence synthesis conducted by \cite{Griffin2020,mcaloon2020incubationBMJOpen,Byrne2020inferredBMJOpen}.

\begin{figure}
\centering
\includegraphics[width=12cm]{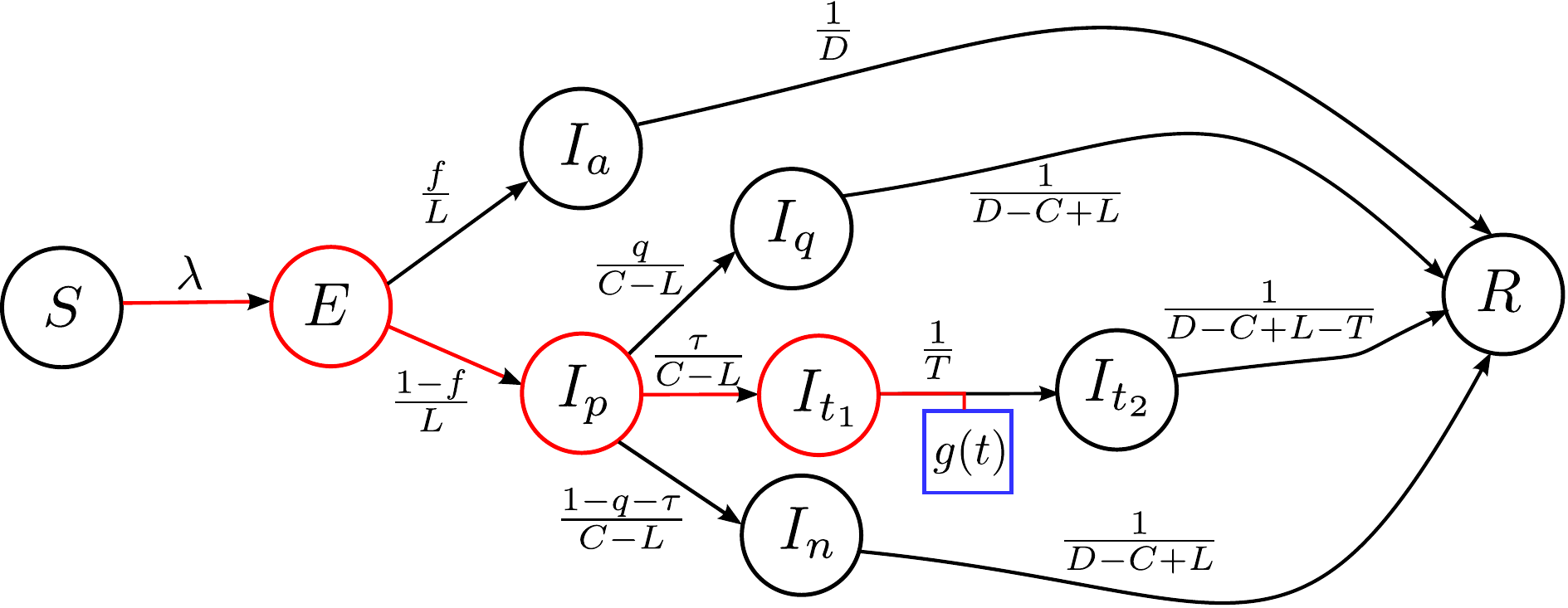}\\
\caption{Representation of the model of Eqs.~(\ref{1})--(\ref{10}) as a weighted, directed network or graph. Each node (vertex) represents a compartment of the model and each edge (link) shows the direction of flow of individuals. The out-edges from a node are weighted by the probability of exiting the node along this edge divided by the average residence time in the compartment of that node. The blue square marks the source of the observed data, fitted by $g(t)$ (see Sec.~\ref{sec:GAM}), which is the daily number of newly-reported confirmed cases. This is related to the flux from $\text{I}_{\text{t}_1}$ to $\text{I}_{\text{t}_2}$ by Eq.~(\ref{10}). The red edges and red nodes are discussed in Conditions~\ref{condition1} and \ref{condition2} of Sec.~\ref{sec:inversion}.}\label{fignetwork}
\end{figure}

The compartments (S, E, etc.) of the model, see Fig.~\ref{fignetwork}, are labelled by the state of the individuals who are assumed to flow through the model structure as their infection progresses. The mathematical variables ($S(t)$, $E(t)$, etc.) represent the number of individuals---from a homogeneous population of fixed size $N$---in each of the respective compartments at time $t$. Those individuals who are in the susceptible (S) compartment can, should they share a contact with an infected individual that enables transmission of the virus, become exposed (enter the E compartment). While individuals are in the exposed compartment (for an average of $L$ days) the virus is still latent so they do not display symptoms, nor are they infectious.

At the end of their latent period, we assume that a fraction $f$ of exposed individuals enter the infected-asymptomatic ($\text{I}_\text{a}$) compartment, where they do not develop symptoms but they may infect others (with a probability of infection that is a factor $h<1$ lower than that of symptomatic infected individuals). The remaining fraction $1-f$ of individuals exiting the exposed compartment flow into the presymptomatic-infected ($\text{I}_\text{p}$) compartment. They remain in the $\text{I}_\text{p}$ compartment for an average duration of $C-L$ days, where $C$ is the incubation period of the virus, and while there they do not show symptoms but they are infectious. Symptoms are assumed to develop at the exit time from the $\text{I}_\text{p}$ compartment and there are three routes that individuals may take: with probability $q$ they self-isolate and quarantine while infectious (in the $\text{I}_\text{q}$ compartment), with probability $\tau$ they undergo a COVID-19 test and isolate while awaiting their result (in the $\text{I}_{\text{t}_1}$ compartment) and the remaining cases (probability $1-q-\tau$) are assumed to not quarantine and remain in the community while infectious ($\text{I}_\text{n}$ compartment). In all cases, the average period of infectiousness is denoted by $D$, while the duration in the presymptomatic compartment is $C-L$ so that, for example, the average time spent in the $\text{I}_\text{q}$ or $\text{I}_\text{n}$ compartments is $D-(C-L)$.

Those individuals who are symptomatic and tested flow from the $\text{I}_{\text{t}_1}$ compartment to the $\text{I}_{\text{t}_2}$ compartment when their test result is confirmed: this occurs an average time $T$ after their symptoms appeared (so the average residence time in the $\text{I}_{\text{t}_1}$ compartment is $T$ and that in the $\text{I}_{\text{t}_2}$ compartment is $D-(C-L)-T$). The main output of the model is the number of new confirmed cases per day, which is the flux from the $\text{I}_{\text{t}_1}$ compartment to the $\text{I}_{\text{t}_2}$ compartment.
After an average duration $D$ of infectiousness, all individuals either recover or are removed (hospitalised or die) and are accounted for in the R compartment.

The flows described above are expressed in terms of differential equations for the time-dependent variables $S(t)$, $E(t)$, etc. as follows:
\begin{align}
\frac{d S}{d t} & = -\lambda S \label{1}\\
\frac{d E}{d t} & =\lambda S - \frac{1}{L}  E\label{2} \\
\frac{d I_a}{d t} & = \frac{f}{L}  E - \frac{1}{D} I_a\label{4} \\
\frac{d I_p}{d t} & = \frac{~(1-f)}{L}  E - \frac{1}{C-L} I_p\label{3} \\
\frac{d I_q}{d t} & = \frac{q}{C-L}  I_p - \frac{1}{D-C+L} I_q\label{5} \\
\frac{d I_{t_1}}{d t} & = \frac{\tau}{C-L}  I_p - \frac{1}{T} I_{t_1}\label{6} \\
\frac{d I_{t_2}}{d t} & =  \frac{1}{T} I_{t_1}-\frac{1}{D-C+L-T} I_{t_2}\label{7}  \\
\frac{d I_{n}}{d t} & =  \frac{(1-q-\tau)}{C-L}  I_p-\frac{1}{D-C+L} I_{n}  \label{8}\\
\frac{d R}{d t} & = \frac{1}{D} I_a + \frac{1}{D-C+L} I_q+\frac{1}{D-C+L-T} I_{t_2}+\frac{1}{D-C+L} I_{n}, \label{9}
\end{align}
where $S(t)$ is the number of susceptible individuals, $E(t)$ is the number who are exposed, $I_p(t)$ is the number who are presymptomatic infected, $I_a(t)$ is the number who are asymptomatic infected, $I_q(t)$ is the number who are symptomatic and self-isolating (without testing), $I_{t_1}(t)$ is the number who are symptomatic and waiting for testing, $I_{t_2}(t)$ is the number who are in post-test self-isolation, $I_n(t)$ is the number who are symptomatic and not isolating and $R(t)$ is the number who are removed (i.e., recovered from the virus or dead). Time $t$ is counted in days from 28th February 2020, to match the timing of the first confirmed cases of COVID-19 in Ireland.

The force of infection $\lambda(t)$ that appears in Eqs.~(\ref{1}) and (\ref{2}) is the time-dependent rate at which susceptible individuals acquire the disease \cite{vynnycky2010introduction}. This is given by the effective contact rate $\beta(t)$ (the total contact rate multiplied by the risk of infection given contact between an infectious and a susceptible person) multiplied by the probability that a contact (in a well-mixed population) is effectively infectious, given by the weighted sum over infectious compartments divided by population size $N$, so $\lambda(t)$ can be written as
\begin{equation}
\lambda(t)=\beta \left(I_p+h I_a+i I_q + I_{t_1}+j I_{t_2}+ I_n\right)/N. \label{forceofinfection}
\end{equation}
Here, the parameters $h$, $i$ and $j$ are multiplicative factors to model the reduction of effective transmission from, respectively, the asymptomatic infected ($\text{I}_{\text{a}}$), symptomatic quarantining  ($\text{I}_{\text{q}}$) and post-test isolation ($\text{I}_{\text{t}_2}$) compartments, relative to symptomatic infected.

In addition, we define $C_c(t)$ to be the cumulative number of new cases reported by time $t$, given by integrating the flux out of the $I_{t_1}$ (waiting-for-test) compartment:
\begin{equation}
\frac{d C_c}{d t} = \frac{1}{T} I_{t_1} \label{10}
\end{equation}
and we also report the number of new daily cases on day $t$, defined by $c_c(t)=C_c(t)-C_c(t-1)$.

The ranges assumed for all parameters are guided by literature reviews and are summarised in the Supplementary Material.

\subsection{Finite-difference formulation} \label{sec:finitediff}
The system of differential equations (\ref{1})--(\ref{10}) may be solved numerically using, for example, a simple forward-Euler scheme. This scheme approximates the derivative $dx/dt$ by $\left(x_{n+1}-x_n\right)/\Delta$, where $\Delta$ is the finite timestep and $x_n$ is the discrete-time approximation to $x(n\, \Delta)$, the value of $x(t)$ at time $t=n\, \Delta$. Expressing the dynamical system (\ref{1})--(\ref{10}) in the general form
\begin{equation}
\frac{d x}{d t} = F\left(x(t)\right),
\end{equation}
where $x(t)$ represents the vector of all unknowns and using the finite-difference approximation enables the solution $x_n$ to be determined from the initial condition $x_0$ by iteration:
\begin{equation}
x_{n+1} = x_n + \Delta \,F\left(x_n\right) \quad\text{ for }n=0,1,2,\ldots.\label{FD2}
\end{equation}
To ensure the accuracy of this finite-difference formulation, it is important that the timestep $\Delta$ chosen be sufficiently small: the convergence of the finite-difference solution to the differential equation solution occurs in the limit $\Delta \to 0$. Testing of the finite-difference results for the system (\ref{1})--(\ref{10}) shows that a value of $\Delta$ equal to 0.1 days gives sufficient accuracy.

\subsection{Network representation}
It proves convenient to represent the model structure described above as a directed weighted graph or network, see Fig.~\ref{fignetwork}. The nodes (vertices) of the graph represent the compartments of the model, while each directed edge shows the direction of flow of individuals as the disease progresses. The out-edges from a node are weighted by the probability of exiting the node along this edge divided by the average residence time in the compartment of that node. Consider, for example, the two out-edges from the E node in Figure~\ref{fignetwork}. The edge leading to the asymptomatic infected compartment (node $\text{I}_{\text{a}}$) has weight $f/L$ because a fraction $f$ of individuals flow along the path while the average time spent in the E compartment is $L$. Similarly, the edge leading to the presymptomatic compartment (node $\text{I}_{\text{p}}$) has weight $(1-f)/L$. The output of the model is the flux (number of individuals per unit time) along the edge from $\text{I}_{\text{t}_1}$ to $\text{I}_{\text{t}_2}$ and this is marked as $g$ in Fig.~\ref{fignetwork}.
We shall show in Sec.~\ref{sec:Calibration} that this directed graph structure facilitates extensions of the model to include more complicated structural features.

Representing the network by its weighted (and time-dependent) adjacency matrix, with $a_{i j}(t)$ being the weight of the directed edge from node $i$ to node $j$ at time $t$ (and $a_{i j}=0$ if there is no edge from node $i$ to node $j$), note that the system of differential equations (\ref{1})--(\ref{10}) is succinctly expressed as
\begin{equation}
    \frac{d x_j}{d t} = \sum_{i} a_{i j} x_i - \sum_k a_{j k} x_j \label{G1},
\end{equation}
where $x_j(t)$ is the number of individuals in compartment $j$ at time $t$. In this equation, the first term on the right-hand side sums over the in-edges to node $j$, while the second term is the sum of the outflows from node $j$. Although this appears to have the form of a linear system, note that the force of infection $\lambda(t)$---which depends on the infectious compartments---is an element of the adjacency matrix, which is therefore time-dependent, and so the system is nonlinear. Nevertheless, the form of Eq.~(\ref{G1}) can be exploited to enable calibration of the model to data.

\section{Calibration}\label{sec:Calibration}
As COVID-19 spread, governments of most countries enacted non-pharmaceutical interventions to slow the growth in the number of cases. These interventions typically aim to reduce the effective contact rate $\beta$ so that there are fewer opportunities for the virus to be transmitted from an infectious to a susceptible person. We therefore assume that the effective contact rate parameter $\beta$ in Eq.~(\ref{forceofinfection}) is time-dependent and we seek to determine what this rate should be in order to reproduce the observed data on the number of confirmed cases. The process we describe here is similar in principle to the method described in \cite{mummert2013studying} (and references therein) for the SIR model, but complicated by the additional compartments of this SEIR model.

\subsection{ Data and Generalized Additive Model}\label{sec:GAM}
The data of interest are the confirmed numbers $c_c(t)$ of COVID-19 cases per day in Ireland. 
The confirmed positive case data was extracted from the Computerised Infectious Disease Reporting (CIDR) database hosted by the Health Protection Surveillance Centre. 
The event date of the cases was used to calibrate the model, with the daily case counts $c_c(t)$ at day $t$ being modelled by a negative binomial random variable
\begin{equation}
 c_c(t) \sim \NB(g(t),\theta), \label{gt0}
\end{equation}
where $g(t)$ is the expected number of cases on day $t$ and $\theta$ is the overdispersion parameter; under this model $\mbox{E}[c_c(t)]=g(t)$ and $\mbox{Var}[c_c(t)]=g(t)+\theta g(t)^2$.

To model the mean parameter $g(t)$ of the negative binomial distribution, we use a thin-plate regression spline\cite{Wood2003},
\begin{equation}
\log g(t) = \beta_0+\sum_{k=1}^{K}\beta_k B_k(t), \label{gt}
\end{equation}
where $(\beta_0, \beta_1, \ldots, \beta_K)$ are unknown parameters and $\{B_k(t):k=1,2,\ldots, K\}$ are thin-plate spline basis functions; the value of $K$ is chosen to be large enough to achieve a satisfactory goodness of fit. The resulting model is a negative binomial generalized additive model (GAM) \cite{Wood2011,Wood2017}. 

To account for parameter uncertainty, Eqs.~(\ref{gt0}) and (\ref{gt}) were fitted in a Bayesian framework using the {\tt brms} R package \cite{Burkner2017,Burkner2018,Burkner2020}. The prior distributions for the model parameters are $\beta_0\sim \mbox{t}_{3}(5.9,2.5)$, $\beta_1,\ldots,\beta_K\sim \mbox{t}_{3}(0,2.5)$ and $\theta\sim\mbox{gamma}(0.01,0.01)$. The {\tt brms} R package interfaces with {\tt Stan} \cite{Stan2020} to generate samples from the posterior distribution for the model parameters.

\subsection{Inversion algorithm}\label{sec:inversion}
The challenge of inverting the SEIR differential equations is the following: for a given set of model parameters and a given fit $g(t)$ of the historical case data, to determine a time-varying effective contact rate $\beta(t)$ and a set of initial conditions for the model of Eqs.~(\ref{1})--(\ref{10}) so that the model output, Eq.~(\ref{10}) exactly matches to the fitted data of Eq.~(\ref{gt}).

Although the steps outlined below can be performed analytically (see Supplementary Material) with a view towards extensions of the model, it is helpful to consider instead the finite-difference approximation of the graph representation given by Eq.~(\ref{G1}). Writing $x_{j,m}$ as the approximation for $x_j(m \Delta )$, the number of individuals in compartment $j$ at timestep $m$, the finite-difference version of Eq.~(\ref{G1}) can be rearranged to give
\begin{equation}
    x_{j,m+1}=x_{j,m}+\Delta\left(\sum_i a_{i j,m}x_{i,m} - \sum_k a_{j k,m} x_{j,m}\right) \text{ for } m=0,1,\ldots, \label{GFD1}
\end{equation}
where we write $a_{i j,m}$ to the denote the $(i,j)$ entry of the time-dependent adjacency matrix at timestep $m$. 
Note that if the time-dependence of all compartments that are in-neighbours of compartment $j$ is known (i.e., if $x_{i,m}$ is known for all $m$ for those compartments $i$ that have $a_{i j, m}>0$), and if an initial condition $x_{j,0}$ is given for compartment $j$, then Eq.~(\ref{GFD1}) provides an explicit iterative scheme to determine $x_{j,m}$ for all $m$. This means that if the time-dependence is known for all in-neighbour nodes of node $j$ then the value of $x_j$ can be fully determined.

A partial converse result also exists. Suppose that the time-dependence of compartment $j$ is known for all time (i.e., $x_{j,m}$ is known for all $m$) and also assume that node $j$ only has one in-neighbour, denoted node $i$. Then the first sum on the right-hand side of Eq.~(\ref{G1}) reduces to a single term, and the corresponding finite-difference approximation can be rearranged to yield
\begin{equation}
    x_{i,m}=\frac{1}{a_{i j,m}}\left(\frac{x_{j,m+1}-x_{j,m}}{\Delta}+\sum_k a_{j k,m} x_{j,m}\right) \text{ for } m = 0,1,\ldots. \label{GFD2}
\end{equation}
Thus, the time dependence of nodes who have a single ``parent'' (in-neighbour) node can, if known, be used to determine the time dependence of the parent node, including the initial condition $x_{i,0}$ of the parent node. 

These two properties---the ability to determine time-dependence of nodes from the time-dependence of all in-neighbours, Eq.~(\ref{GFD1}), and the partial converse of determining the time-dependence of a single parent node from that of its ``child'' node, Eq.~(\ref{GFD2})---form the basis of the algorithm that allows calibration of the model to output data from a range of graph-represented models.

The first step of the algorithm is to link the model's time-dependent output directly to the time dependence of a compartment (node). In all our models the output can be expressed in terms of the cumulative number of confirmed cases $C_c(t)$ and if this is given (e.g., by the GAM fit of Eq.~(\ref{gt})) so that $d C_c/dt =g(t)$ is a known function, then Eq.~(\ref{10}) can be inverted to determine $I_{t_1}(t)$ as
\begin{equation}
    I_{t_1}=T g(t) \label{GFD3}.
\end{equation}
Considering the graph representation in Fig.~\ref{fignetwork}, we note that node $\text{I}_{\text{t}_1}$ has a single in-neighbour or parent node, $\text{I}_{\text{p}}$  and so Eq.~(\ref{GFD2}) can be employed to determine the time-dependence of compartment $\text{I}_{\text{p}}$ from the known (from Eq.~(\ref{GFD3})) time-dependence of $\text{I}_{\text{t}_1}$. Similarly, $\text{I}_{\text{p}}$ has a single parent node, E, whose time-dependence can be determined from another application of Eq.~(\ref{GFD2}). These steps of the algorithm---from a node to its single parent---are marked with red edges in Fig.~\ref{fignetwork} and these link the time-dependence of nodes $\text{I}_{\text{t}_1}$, $\text{I}_{\text{p}}$ and E directly to the data-fitting function $g(t)$.

Recalling from Eq.~(\ref{forceofinfection}) that the force of infection $\lambda(t)$ is a nonlinear function of the infectious compartments, some of which have (as yet) unknown time-dependence, the calculation for $S(t)$ is slightly more involved. It proves convenient to define an auxiliary variable $\omega(t)$ as $\omega(t)=\lambda(t)S(t)$, so that Eq.~(\ref{2}) can be rearranged to give
\begin{equation}
    \omega = \frac{d E}{d t}+ \frac{1}{L} E . \label{GFD4}
\end{equation}
Since we have already determined the time-dependence of $E(t)$, at least through its finite-difference approximation, this relationship gives us the time-dependence of the auxiliary variable $\omega(t)$. Then, noting that Eq.~(\ref{1}) is $\frac{d S}{d t} = -\omega$, we can determine the time-dependence of the susceptible compartment from its initial disease-free condition $S(0)=N$ by integration of $\omega(t)$ or, in the finite-difference approximation, by summation. 

We also need to determine the time-dependence of those infectious compartments that are not already marked as red in Fig.~\ref{fignetwork}. The tree-like structure of the infectious nodes, with the E compartment as the ``root'' of the tree, means that the time-dependence of all the nodes can be determined by repeated application of Eq.~(\ref{GFD1}) along with the assumption that the initial condition for any of the ``children'' nodes is zero. In this way, the known time-dependence of variable $E(t)$ determines $I_a(t)$, while knowing $I_p(t)$ allows for $I_q(t)$ and $I_n(t)$ to be determined. Finally, $I_{t_2}(t)$ can be determined from $I_{t_1}$ in the same way.

At this stage in the algorithm we have determined the time-dependence of all infectious compartments and of the auxiliary variable $\omega(t)$. We can therefore rearrange Eq.~(\ref{1}) to explicitly solve for the effective contact rate $\beta(t)$ as 
\begin{align}
\beta(t) &= - \frac{N}{S} \frac{d S}{d t} \left(I_p+h I_a+i I_q + I_{t_1}+j I_{t_2}+ I_n\right)^{-1} \label{betainfer0}\\
&= \frac{N \omega}{S}  \left(I_p+h I_a+i I_q + I_{t_1}+j I_{t_2}+ I_n\right)^{-1}.\label{betainfer}
\end{align}

Although all the steps outlined above can also be performed analytically by solving algebraic and  linear differential equations (see Supplementary Material), the finite-difference formulation is particularly useful when considering the generalization of the method to other models that are represented by larger graphs. Therefore, it is worth pausing here to consider the three crucial aspects of the graph structure that are exploited in the algorithm. We express these as three conditions for the method to be applicable. In order to do so, we define the concepts of ``red path'' and ``red nodes'', reflecting the colours used for edges and nodes, respectively, in Fig.~\ref{fignetwork}.
\begin{condition} \label{condition1} A reverse-direction ``red path'' exists from the edge of the data-fitting function $g(t)$  to the force-of-infection edge (the edge from S to E that is weighted by $\lambda$), with each node on this red path having exactly one in-neighbour.
\end{condition}
The links that constitute the red path for the IEMAG model are coloured red in Fig.~\ref{fignetwork}. 
Next, we iteratively define the set of ``red nodes'' by beginning with all nodes that are on the red path (these nodes are coloured red in Fig.~\ref{fignetwork}). Then any node whose in-neighbours are all red nodes is also labelled as a red node. We continue this iteration until no further nodes become red nodes, and then the following condition must hold for the calibration algorithm to apply:
\begin{condition} \label{condition2} All infectious-compartment nodes are red nodes.
\end{condition}
A third condition is that the data-fitting function $g(t)$ must be sufficiently smooth. We show in the Supplementary Material that the following condition is necessary:
\begin{condition} \label{condition3} The function $g(t)$ must have at least as many derivatives as the number of edges in the red path.
\end{condition}
Condition~\ref{condition1} enables the identification of the time-dependence of the red-path nodes from the data by application of Eq.~(\ref{GFD2}), while Condition~\ref{condition2} subsequently allows all infectious compartments to be calculated (assuming zero initial condition) from the time-dependence of the red-path nodes using Eq.~(\ref{GFD1}). Once these time dependencies are calculated, the finite-difference values for the auxiliary variable $\omega(t)=\lambda(t)S(t)$ and subsequently for the effective contact rate $\beta(t)$ are evaluated similar to Eq.~(\ref{betainfer}). Condition~\ref{condition3} is required because each step on the red path from $g(t)$ back to S involves differentiating the ``child'' variable to obtain the ``parent'' (in-neighbour) variable, so $g(t)$ must be sufficiently smooth to allow this to occur for all steps on the red path. This has implications for extending the model to allow for compartmental residence-time distributions that are Erlang rather than exponential, see Supplementary Material.



\section{Examples}\label{sec:examples}
To demonstrate the calibration algorithm of Sec.~\ref{sec:inversion} and illustrate its application to scenario analysis, we use the confirmed number of COVID-19 cases per day in Ireland from 28th February 2020 to 11th November 2020, as described in Sec.~\ref{sec:GAM}. The role of IEMAG was to advise on policy decisions, for which various future scenarios were modelled. Given the evolving information on the virus parameters and the noisy nature of daily case data, it is important to quantify the uncertainty associated with any scenario prediction. Accordingly, for each of Figs.~\ref{figRt} and \ref{figscenario2} a sample of 1000 posterior realizations of the curves are generated from the GAM fitting procedure (Sec.~\ref{sec:GAM}), and for each such realization, independent draws from the distributions of the model parameters (see Supplementary Material) are used within the calibration algorithm to generate scenario outputs. 

\subsection{Model-inferred reproduction number}\label{sec:R}
The calibration algorithm described in Sec.~\ref{sec:Calibration} yields the time-dependent effective contact rate $\beta(t)$ that is consistent with the GAM fit $g(t)$ to the case data and with the set of model parameters selected in that realization. It is useful to communicate the values of $\beta(t)$ in terms of a quantity that is closely related to a (time-varying) reproduction number. The basic reproduction number $R_0$ for compartmental models of the type studied here is known to be directly proportional to the effective contact rate $\beta$ through the eigenvalue of a matrix  (see Supplementary Material). This allows us to express effective contact rates $\beta$ in terms of ``equivalent'' $R_0$ values, although care must be taken with interpreting such values as $R_0$ values are only well-defined at the beginning of an epidemic. We call these ``model-inferred $R$'' values. Using this relationship, the time-varying $\beta(t)$ can similarly be expressed as a time-varying $R$ value, and the $\beta$ values considered in the scenario analyses below are described in terms of their equivalent model-inferred $R$ values.

\begin{figure}
\centering
\includegraphics[width=7cm]{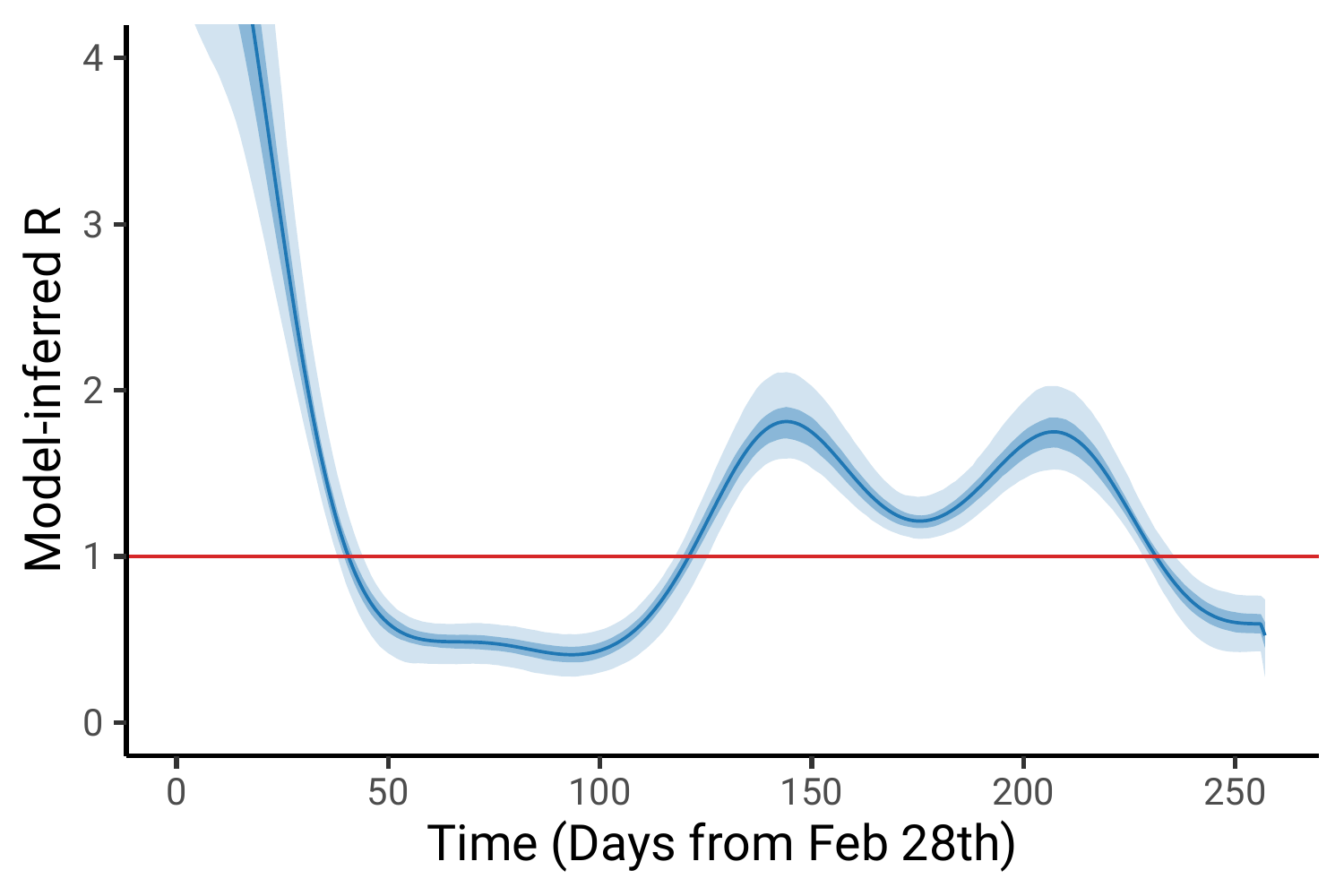}\\
\caption{Model-inferred $R$ up to 11th November 2020 (day 257 of the Irish epidemic), as described in Section~\ref{sec:examples}\ref{sec:R}. The curve indicates the mean (over $1000$ realizations as described in Sec.~\ref{sec:examples});  
shaded regions show the $50\%$ (quantiles $0.25$ to $0.75$) and $95\%$ (quantiles $0.025$ to $0.975$) credible intervals. In this and subsequent figures time is measured in days from February 28th, 2020.}\label{figRt}
\end{figure}

Figure~\ref{figRt} shows the model-inferred $R$ (which is directly proportional to $\beta(t)$) determined by the calibration algorithm on 11th November 2020 (day 257), as a function of time in terms of days from 28th February 2020. The periods where $R$ is below one correspond to declining case numbers in the population. While the times where $R$ changes from above to below one, and vice versa, can be related to the days when population-level movement restrictions changed, they are not exactly the same, which demonstrates the utility of a data-driven approach to estimation rather than specifying  breakpoints where $\beta(t)$ is assumed to change value.

\subsection{Scenarios}\label{sec:scenarios}
For each set of model parameters, the calibration algorithm detailed in Section~\ref{sec:Calibration} yields a time-varying contact rate $\beta(t)$ and a set of initial conditions that are consistent with the GAM-fitted data up to the calibration date of 11th November 2020. Forward predictions of the model from that date may then be examined under assumptions about how the contact rate (or, equivalently, the model-inferred reproduction number) may behave in the future.

\begin{figure}
\hspace{-0.5cm}
\includegraphics[width=7cm]{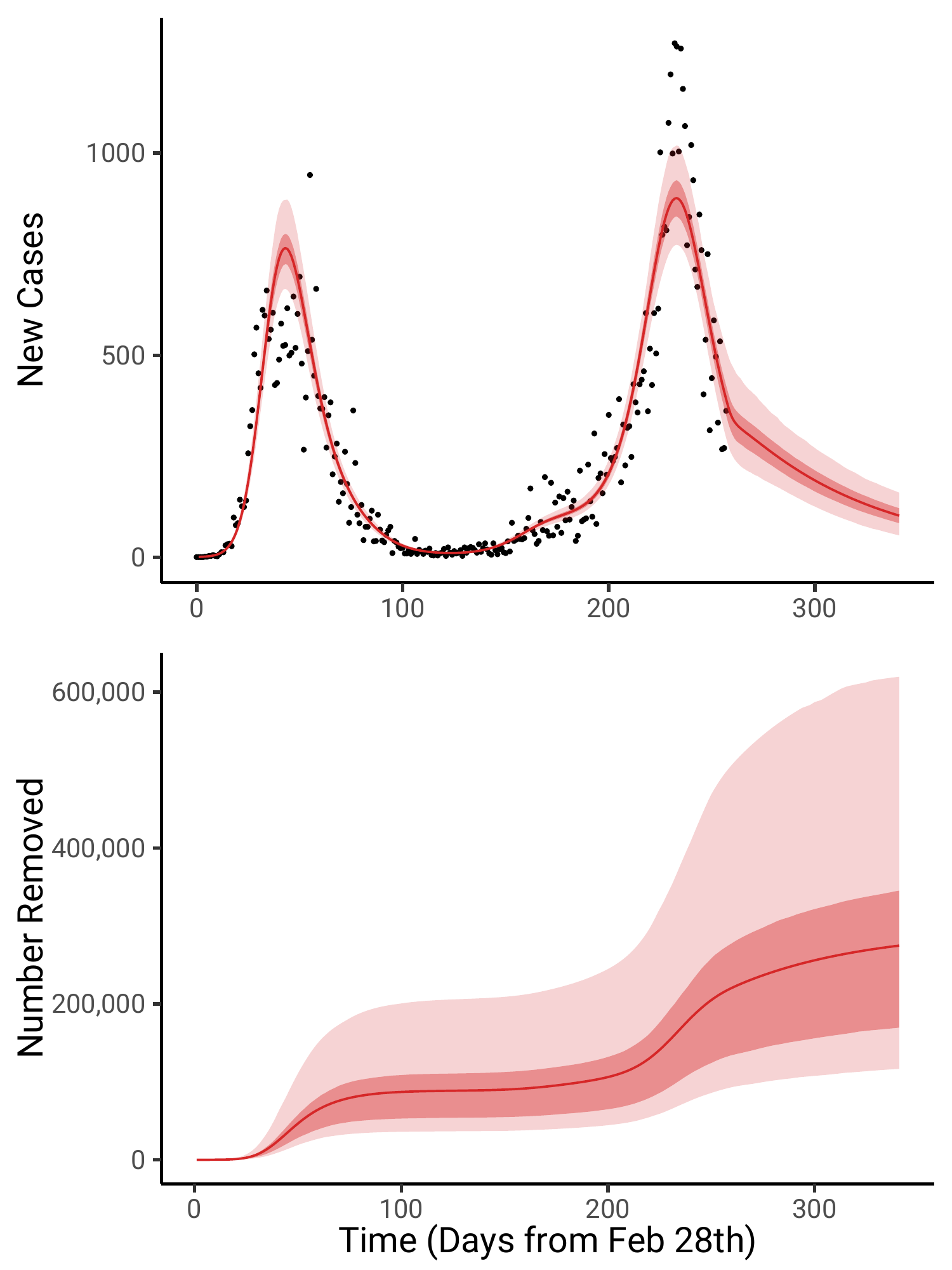}\includegraphics[width=7cm]{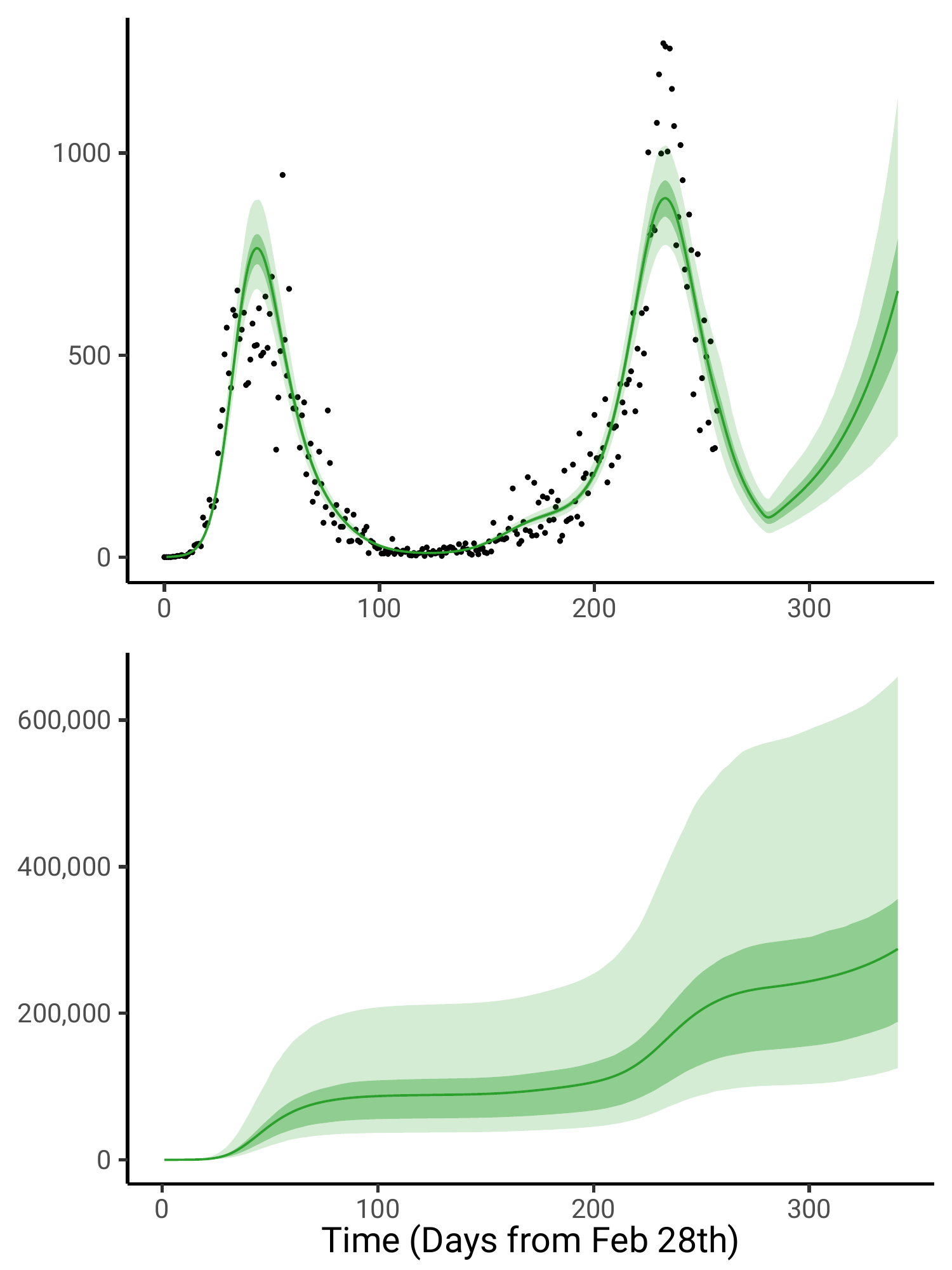}\\
\caption{Scenario 1 (left panels) assumes $R=0.9$ from 11th November (day 257); scenario 2 (right panels) assumes $R=0.5$ from 11th November until 2nd December (day 278), with $R=1.4$ thereafter. The top panels show confirmed new cases per day, with black symbols for the observed data; the bottom panels show the number of removed individuals (this includes both deaths and those who recovered from the virus). In each case the curve indicates the mean (over $1000$ realizations as described in Sec.~\ref{sec:examples});  shaded regions show the $50\%$ (quantiles $0.25$ to $0.75$) and $95\%$ (quantiles $0.025$ to $0.975$)  credible  intervals.}\label{figscenario2}
\end{figure}

As an example, in Scenario 1, shown in the left panels of Figure~\ref{figscenario2}, we assume that the effective contact rate will remain at a level equivalent to $R=0.9$ from the calibration date. Solving the finite-difference approximation (Eq.~\ref{FD2}) of the differential equation system given by Eqs.~(\ref{1})--(\ref{10}) yields the expected number of daily cases and other outputs, such as the number removed (number of individuals in the R compartment) under this assumption, see Fig.~\ref{figscenario2}.

Scenario 2, in the right panels of Fig.~\ref{figscenario2}, investigates an alternative possibility for the post-calibration-date contact rates. Here, the inferred reproduction number $R$ is assumed to be 0.5 from 11th November (day 257) until 2nd December (day 278), with $R$ increased to 1.4 thereafter. It is  important to note that in each case it is assumed that the reproduction number is exactly as specified in the scenario description. Comparing the results for the two scenarios, it is clear that the effect of uncertainty in future $R$ values dominates the uncertainty due to the range of possible values for the other parameters of the model.

\subsection{Incorporating vaccination into the model}
As population-level vaccination is an important instrument for virus suppression, it is important to extend the model described in Sec.~\ref{sec:IEMAGSEIR2} to include the impact of policy decisions related to vaccine rollout. 

\begin{figure}
\centering
\includegraphics[width=12cm]{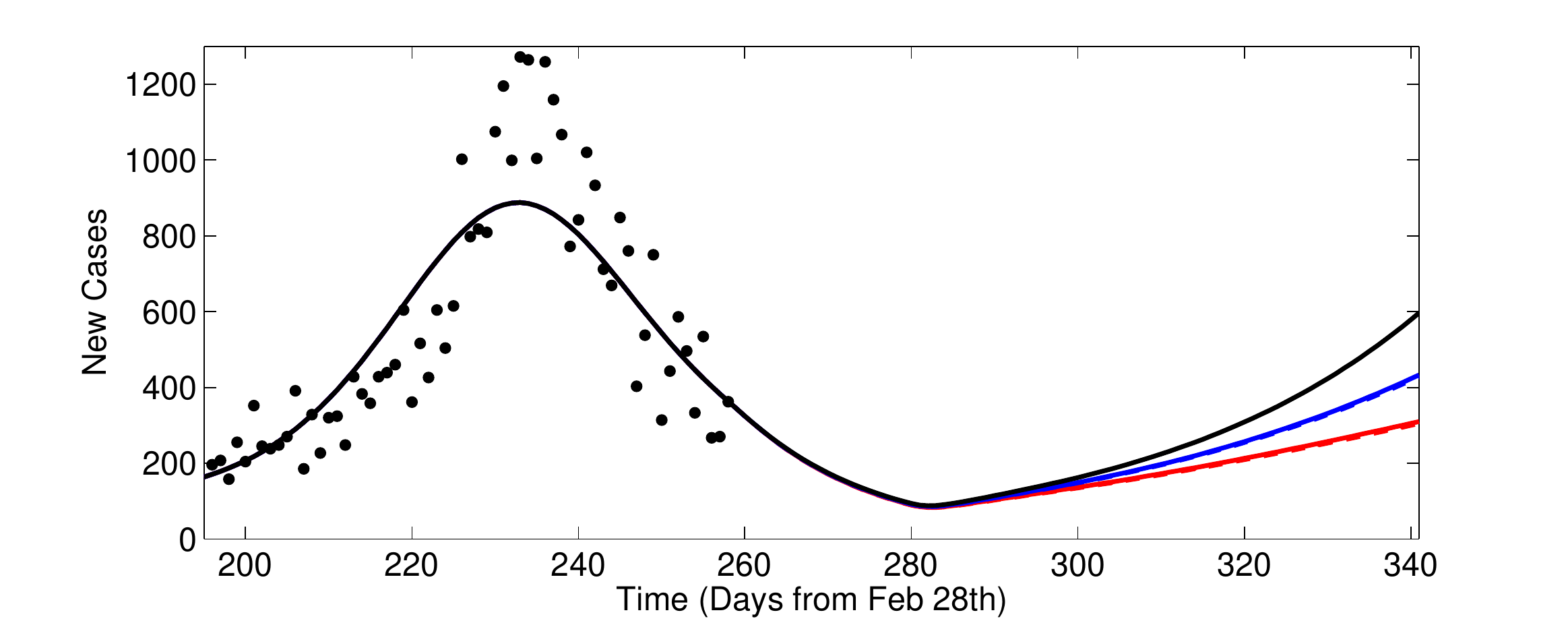}\\
\caption{Daily new cases for a single realization of parameter values (see Supplementary Material), assuming $R=0.5$ from 11th November (day 257) until 2nd December (day 278), with $R=1.4$ thereafter. The black curve is from the model of Eqs.~(\ref{1})--(\ref{10}), with no vaccinations. The blue and red solid curves are for scenarios from the vaccination model described in Supplementary Material, which assume vaccines are administered at a rate of $v_d=5000$ per day (blue) or $v_d=10000$ per day (red), with the vaccines first becoming effective on 11th November. Almost indistinguishable from the solid curves are the blue and red dashed curves, which show the results of the reduced-order model described in Supplementary Material for $v_d=5000$ and $v_d=10000$, respectively.  }\label{figvacc}
\end{figure}

In the Supplementary Material we describe a model that includes new compartments called SV, EV, IV and RV to contain those individuals who are effectively vaccinated while also being susceptible, exposed, (asymptomatic) infected or removed, respectively. Figure~\ref{figvacc} shows how vaccinations, assumed to be administered at a constant rate of $v_d$ per day, could impact on Scenario 2. For clarity we show results only for one set of parameters (see Supplementary Material for parameter values) but we illustrate the impact of the vaccination rollout speed by comparing results for scenarios where we assume either $v_d=5000$ or $v_d=10000$ vaccines are administered each day, becoming effective from 11th November (day 257). The calibration and scenario match that of Scenario 2, with an effective contact rate equivalent to $R=0.5$ switching to $R=1.4$ on 2nd December (day 278). 

For further extensions of the model, for example including age-cohort compartments for both vaccinated and unvaccinated individuals, it proves convenient to derive a reduced-order approximation to the vaccination model. The derivation of this reduced model is described in the Supplementary Material. The results of the reduced model are shown by the dashed curves in Fig.~\ref{figvacc} and they are almost indistinguishable from the full model results (solid curves), demonstrating the accuracy of the reduced model.

\section{Conclusion}\label{sec:conclusion}
In this paper we have outlined the population-level SEIR model of COVID-19 that is used by the Irish Epidemiological Modelling Advisory Group (IEMAG). Our main result is the calibration algorithm of Sec.~\ref{sec:Calibration} that is sufficiently general to allow for extensions to the model structure. Our analysis identified three conditions on the graph structure of the model and the smoothness of the data-fitting function that are necessary and sufficient for the inversion algorithm to be applicable. We gave examples in Sec.~\ref{sec:examples} of the use of the model for scenario analysis and uncertainty quantification; we also demonstrated the adaptability of the calibration algorithm by applying it to a model with vaccination. While vaccination requires a substantial increase in the number of model compartments, we showed that an approximate reduced-order model can also give good accuracy at lower complexity.
 
 There are several directions in which this work could be advanced. Our negative binomial GAM fits (Eq.~(\ref{gt0})), for example, assume a constant overdispersion parameter but this could be generalised to allow for time-dependent overdispersion. Another possible extension would be to consider multiple data sources as required, for example, in an age-cohorted SEIR model where daily confirmed cases are recorded for various age categories, or to include time series data on hospitalizations, deaths or multi-strain variants in addition to case-counts.
 We hope that the work presented here will form a basis for such extensions and that our focus on network-representation calibration will facilitate adoption of the approach for  other SEIR-type pandemic models.




\section*{Code}
R code that implements the calibration algorithm described in Sec.~\ref{sec:Calibration} and runs an example scenario is available for download from \url{https://github.com/obrienjoey/ireland_covid_modelling}.



\section*{Acknowledgements}
The models were constructed with the advice of members of the Irish Epidemiological Modelling Advisory Group.
This work was partly funded by Science Foundation Ireland  through grant numbers SFI/16/IA/4470 (JG),  SFI/12/RC/2289\_P2 (JG,TBM,NF,NB), SFI/16/RC/3918 (JG,NB),  SFI/12/RC/2275\_P2 (NB), and  SFI/16/RC/3835 (TBM).



\bibliographystyle{unsrt}
\bibliography{covid_refs}









\clearpage
\section*{Supplementary Material}
\beginsupplement
\setcounter{section}{1}
 \renewcommand{\thesection}{SM\arabic{section}}
 \renewcommand{\thesubsection}{SM\arabic{subsection}}
 \setcounter{equation}{0}
 \renewcommand{\theequation}{S\arabic{equation}}

In this Supplementary Material file we give further details of the models and parameters used in the main text. Section~\ref{SMa} describes the ranges of model parameter used to quantify uncertainty in Figs.~\ref{figRt} and \ref{figscenario2}. Section~\ref{SMb} gives the details of the relation between the effective contact rate $\beta$ and the basic reproduction number $R_0$. Section~\ref{SMc} discusses the possibility of modelling Erlang residence-time distributions for the compartments instead of exponential distributions. In Section~\ref{SMd} we give details of the analytical solution to the calibration inversion algorithm of Sec.~\ref{sec:inversion}. Sections~\ref{SMe} and \ref{SMf} give details of the vaccination model and the reduced-order vaccination model that are used in Fig.~\ref{figvacc}.

\subsection{Ranges of parameter values} \label{SMa}
For the scenarios shown in Figs.~\ref{figRt} and \ref{figscenario2}, in every realization each model parameter is independently chosen from a uniform distribution with upper and lower limits as listed below. (All timescales are expressed in days. Day 0 of the Irish epidemic is February 28, 2020.)
\begin{itemize}
\item $L$: average latent period; range assumed is $3.9$ to $5.9$.
\item $C$: average incubation period; range assumed is $\max(L,4.8)$ to $6.8$ (lower limit ensures $C-L$ is nonnegative).
\item $D$: average infectious period; range assumed is  $\max(C-L,5.0)$ to $9.0$ (lower limit ensures $D-(C-L)$ is nonnegative).
\item $h$: multiplicative factor for reduction of effective transmission from the asymptomatic infected compartment $\text{I}_\text{a}$, relative to symptomatic infected; range assumed is $0.01$ to $0.5$.
\item $i$: multiplicative factor for reduction of effective transmission from the self-quarantine compartment $\text{I}_\text{q}$, relative to symptomatic infected; range assumed is $0$ to $0.1$.
\item $j$: multiplicative factor for reduction of effective transmission from the post-test isolation compartment $\text{I}_{\text{t}_2}$, relative to symptomatic infected; range assumed is $0$ to $0.1$.
\item $f$: fraction of infected who are asymptomatic; range assumed is $0.18$ to $0.82$.
\item $\tau$: fraction of symptomatic cases that are tested; range assumed is $0.5$ to $1.0$.
\item $q$: fraction of symptomatic cases that self-quarantine from appearance of symptoms until recovery; range assumed is $0$ to $1-\tau$ (upper limit ensures $1-q-\tau$ is nonnegative).
\item $T$: average wait for test results; range assumed is $1.0$ to $\max(5.0, D-C+L)$ (upper limit ensures $D-C+L-T$ is nonnegative).
\item $N$: population, assumed to be $4.9\times 10^6$.
\end{itemize}
As described in Sec.~\ref{sec:Calibration}, the effective contact rate $\beta$ is assumed to be time-dependent, in order to model the impact of interventions. 

For the single-realization case of Fig.~\ref{figvacc}, the parameters  were set to their approximate mid-range values, as follows: $L=4.9$, $C=5.9$, $D=7.0$, $h=0.25$, $i=0.05$, $j=0.05$, $f=0.5$, $\tau=0.75$, $q=0.13$, and $T=3.6$. The vaccination parameters were set to $\epsilon=0.8$, $f_2=0.5$ and $h_2=0.125$.


\subsection{Basic reproduction number} \label{SMb}
To calculate the basic reproduction number $R_0$ for the model we follow the approach explained in Section 2.2 of \cite{Heffernan05}. The value of $R_0$ is given by the spectral radius of the next generation matrix, which can be written as $F V^{-1}$ for certain matrices $F$ and $V$. Defining the vector of relevant dynamical variables as $\mathbf{x}=\{ E, I_p, I_a, I_q, I_{t_1}, I_{t_2}, I_n\}$, the matrix $F$ is zero except for its first row, which is $(0,\beta,h \beta,i \beta,\beta,j \beta,\beta)$. The corresponding matrix $V$ is
\begin{equation}
\left(
\begin{array}{ccccccc}
 \frac{1}{L} & 0 & 0 & 0 & 0 & 0 & 0 \\
 -\frac{(1-f)}{L} & \frac{1}{C-L} & 0 & 0 & 0 & 0 & 0 \\
  -\frac{f}{L} & 0 & \frac{1}{D} & 0 & 0 & 0 & 0 \\
 0 & -\frac{ q}{C-L} & 0 & \frac{1}{D-C+L} & 0 & 0 & 0 \\
 0 & -\frac{ \tau}{C-L} & 0 & 0 & \frac{1}{T} & 0 & 0 \\
 0 & 0 & 0 & 0 & -\frac{1}{T} & \frac{1}{D-C+L-T} & 0 \\
 0 & -\frac{ (1-q-\tau)}{C-L} & 0 & 0 & 0 & 0 &
   \frac{1}{D-C+L} \\
\end{array}
\right).
\end{equation}
Calculating the eigenvalues of $F V^{-1}$ then yields the relationship that we use to express $\beta$ values in terms of equivalent model-inferred $R$ values, see  Sec.~\ref{sec:R}:
\begin{align}
\frac{R_0}{\beta}& = (f-1) ((i-1) q (C-L)+(j-1) \tau (C-L+T))+\nonumber\\
& +D (f (h-i q-j \tau+q+\tau-1)+(i-1) q+(j-1) \tau+1).
\label{R0formula}
\end{align}


It can prove useful to also consider this formula in terms of contributions from each of the infected compartments. With a contribution to $R_0$ defined as the fraction of infectives moving through the compartment multiplied by the average time spend in the compartment multiplied by the transmission rate for that compartment, the respective contribution from each compartment is as follows:
\begin{itemize}
\item Contribution from $I_a$ is $f D h \beta$.
\item Contribution from $I_p$ is $(1-f) (C-L) \beta$.
\item Contribution from $I_q$ is $(1-f)q (D-C+L) i \beta$.
\item Contribution from $I_{t_1}$ is $(1-f) \tau T \beta$.
\item Contribution from $I_{t_2}$ is $(1-f)\tau (D-C+L-T) j \beta$.
\item Contribution from $I_n$ is $(1-f)(1-q-\tau)(D-C+L) \beta$.
\end{itemize}
Summing these contributions gives Equation~(\ref{R0formula}).

\subsection{Erlang residence-time distributions}\label{SMc}
Another possible application of the graph-based representation introduced in the main text is to allow each of the exposed and infectious compartments to be replaced by $n$ sub-compartments. The derivation of the deterministic equations of Sec.~\ref{sec:IEMAGSEIR2} assumes that an individual spends an exponentially-distributed time in any one compartment. By replacing, for example, the $E$ compartment which has exit rate $1/L$ by a string of $n$ sub-compartments, each with exit rate $n/L$, the overall time spent in the exposed sub-compartments has an Erlang-$n$ distribution, which can better approximate the observed distributions for the various stages of the disease. However, since adding sub-compartments leads to a longer graph path from the data-fit $g(t)$ to the susceptible node in the first step of the calibration algorithm, from Condition~\ref{condition3} the function $g(t)$ in Eq.~(\ref{gt}) would be required to have higher levels of smoothness than the thin-plate spline fit used here. Accordingly, we limited this study only to compartments with exponentially-distributed residence times.

\subsection{Analytical results for model calibration}\label{SMd}
In this section we solve a series of linear differential equations and algebraic equations to determine $\beta(t)$ analytically for the model of Eqs.~(\ref{1})--(\ref{10}): this procedure is the continuous-time analogue of the finite-difference approach described in Sec.~\ref{sec:inversion}.

Define the function fitting the daily-count data as $g(t)$, and assume that
\begin{equation}
g(0)=0.
\end{equation}
Equation~(\ref{GFD3}) allows us to immediately infer $I_{t_1}(t)$. We can then rearrange Eq.~(\ref{6}) to express $I_p(t)$ in terms of $I_{t_1}(t)$ and its derivative (i.e., we are taking the first step upon the red path defined in Condition~\ref{condition1}). Knowing $I_p(t)$, we can similarly rearrange Eq.~(\ref{3}) to determine $E(t)$ in terms of $I_p(t)$ and its derivative. The next step is to calculate the auxiliary variable $\omega(t)$ from the known time-dependence of $E(t)$: this is done in Eq.~(\ref{GFD4}). Once we know $\omega(t)$ we determine $S(t)$ from the equation $dS/dt=-\omega$ by integration, using the initial condition $S(0)=N$.

We then move to calculate the time-dependence of the remaining infectious compartments from the ``red nodes'' whose time-dependence is now known. Equation~(\ref{4}) is a linear differential equation for $I_a(t)$ that has the known function $E(t)$ as a forcing term; assuming that $I_a(0)=0$, the equation can be solved by, for example, Laplace transform techniques. Similarly, $I_q(t)$ and $I_n(t)$ are determined from the known $I_p(t)$ by solving the linear differential equations~(\ref{5}) and (\ref{8}), respectively, with zero initial conditions. Finally, $I_{t_2}(t)$ is determined from $I_{t_1}(t)$ by solving the linear differential equation~(\ref{7}) with $I_{t_2}(0)=0$.


As a result of this procedure, the variables that appear in Eq.~(\ref{betainfer0}) can be written explicitly in terms of the data-fitting function $g(t)$ as follows 
\begin{align}
S(t) & = c_{1 1}+c_{1 2}\,g'(0)+c_{1 3}\,\int_0^t g(s)\, ds + c_{1 4}\, g(t) + c_{1 5}\,g'(t)+c_{1 6} \, g''(t),\label{anal1}\\
S'(t) & = c_{1 3}\, g(t) + c_{1 4}\, g'(t) + c_{1 5}\,g''(t) + c_{1 6}\,g'''(t),\label{Sprime}\\
I_p(t) & = c_{2 1}\, g(t) + c_{2 2}\, g'(t),\\
I_a(t) & = c_{3 1}\, \int_0^t e^{-k_1(t-s)}g(s)\, ds + c_{3 2}\, g(t) + c_{3 3}\, g'(t) + c_{3 4}\,g'(0) e^{-k_1 t},\\
I_i(t) &= c_{4 1}\, \int_0^t e^{-k_2(t-s)}g(s)\, ds  + c_{4 2}\, g(t),\\
I_{{t_1}}(t) &= c_{5 1} \,g(t),\\
I_{t_2}(t) &= c_{6 1}\, \int_0^t e^{-k_3(t-s)}g(s)\, ds, \\
I_n(t) &= c_{7 1}\, \int_0^t e^{-k_2(t-s)}g(s)\, ds + c_{7 2}\, g(t),\label{analend}
\end{align}
where primes denote derivatives and the coefficients $c_{i j}$ and rates $k_i$ depend on the model parameters (but not on $g(t)$) and are listed in Table~\ref{tab:coeffs}. Note that  $g'''(t)$ appears in Eq.~(\ref{Sprime}) and so the fitting function must have at least three derivatives. This requirement can be traced back to the three ``red-path'' steps (marked in Fig.~\ref{fignetwork}) taken to determine $I_p(t)$ from $I_{t_1}(t)$, then $E(t)$ from $I_p(t)$ and finally $\omega(t)$ from $E(t)$: each step requires one further derivative of the input variable and since $I_{t_1}(t)$ is proportional to $g(t)$, this implies three derivatives of $g(t)$ are required. In more complex models with longer red paths, a similar argument leads to the conclusion expressed as Condition~\ref{condition3} in Sec.~\ref{sec:inversion}.

\begin{table}[!h]
\caption{Coefficients and rates for the analytical solutions of Eqs.~(\ref{anal1})--(\ref{analend})}
\label{tab:coeffs}
\begin{tabular}{llll}
\hline
coefficient & formula \\
\hline
$c_{11}$ & $N$ \\
$c_{12}$ & $\frac{f(C-L) T}{\tau(1-f)}$\\
$c_{13}$ & $\frac{-1}{\tau(1-f)}$\\
$c_{14}$ & $\frac{-(C+T)}{\tau(1-f)}$\\
$c_{15}$ & $\frac{L^2-C(L+T)}{\tau(1-f)}$\\
$c_{16}$ & $\frac{-(C-L)L T}{\tau(1-f)}$\\
$c_{21}$ & $\frac{C-L}{\tau}$\\
$c_{22}$ & $\frac{(C-L)T}{\tau}$\\
$c_{31}$ & $\frac{f(D-C+L)(D-T)}{D^2(1-f)\tau}$\\
$c_{32}$ & $\frac{f( (D+L) T +C(D-T)-D L)}{D(1-f)\tau}$\\
$c_{33}$ & $\frac{f(C-L)T}{\tau(1-f)}$\\
$c_{34}$ & $\frac{-f(C-L)T}{\tau(1-f)}$\\
$c_{41}$ & $\frac{q T}{\tau}\left(\frac{1}{T}-\frac{1}{D-C+L}\right)$\\
$c_{42}$ & $\frac{q T}{\tau}$\\
$c_{51}$ & $T$ \\
$c_{61}$ & $1$\\
$c_{71}$ & $\frac{T(1-q-\tau)}{\tau}\left(\frac{1}{T}-\frac{1}{D-C+L}\right)$\\
$c_{72}$ & $\frac{T(1-q-\tau)}{\tau}$\\
$k_1$ & $\frac{1}{D}$\\
$k_2$ & $\frac{1}{D-C+L}$\\
$k_3$ & $\frac{1}{D-C+L-T}$
 \\\hline
\end{tabular}
\vspace*{-4pt}
\end{table}

\subsection{Vaccination model}\label{SMe}
\begin{figure}
\centering
\includegraphics[width=12cm]{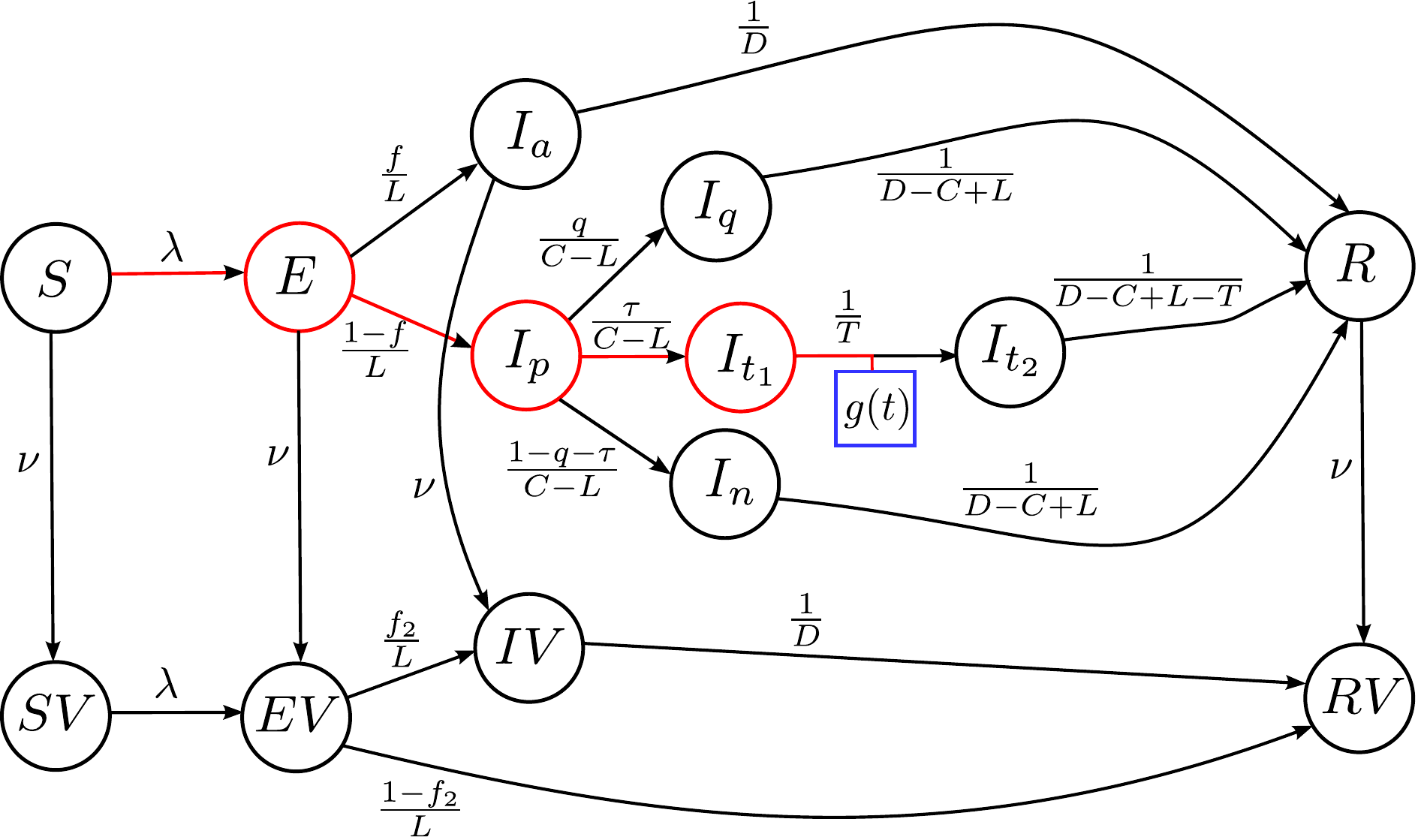}\\
\caption{Extension of Fig.~\ref{fignetwork} to include compartments for effectively-vaccinated individuals (SV, EV, IV and RV). The rate that individuals move from an unvaccinated compartment (e.g., S) to its effectively-vaccinated analogue (e.g., SV) is given in terms of the number of vaccines administered daily by Eq.~(\ref{nu}).}\label{figschematic2}
\end{figure}

In this section we describe an extension to the original model of Fig.~\ref{fignetwork} to include compartments for individuals who are effectively vaccinated while also being susceptible, exposed, (asymptomatic) infected or removed. These compartments are called SV, EV, IV and RV, respectively; see Fig.~\ref{figschematic2}.

By ``effectively vaccinated'' we mean that the individual has received a vaccine shot and that the vaccine is effective for them, meaning that it prevents symptomatic infection. The current understanding is that COVID-19 vaccines are highly effective at preventing symptomatic disease but the impact of the vaccine on the possible transmission from a vaccinated but (asymptomatic) infected individual in the IV compartment has not yet been reliably quantified. Thus, our model assumes that a fraction $\epsilon=0.8$ of individuals who receive the vaccine become effectively vaccinated; the remaining fraction $1-\epsilon=0.2$ are modelled as if they were unvaccinated. Those who are effectively vaccinated and exposed to the virus may become asymptomatic infected---a fraction $f_2$ of individuals who are in the EV compartment flow to the IV compartment---or their virus load may remain so low that we treat them as removed (immune) individuals, so the remaining fraction $1-f_2$ of individuals exiting the EV compartment flow directly to the RV compartment, see the bottom half of Fig.~\ref{figschematic2}.

Those individuals who are in the IV compartment contribute to the force of infection with a multiplicative factor $h_2$ to model the reduced level of effective transmission relative to symptomatic infected, so we update Eq.~(\ref{forceofinfection}) to include an extra term:
\begin{equation}
\lambda(t)=\beta \left(I_p+h I_a+i I_q + I_{t_1}+j I_{t_2}+ I_n+h_2 IV\right)/N. \label{forceofinfection2}
\end{equation}

The rate of flow from a compartment such as S into its effectively-vaccinated version SV is specified in the terms of the expected number of vaccines that become effective at each timestep. Given a population-level vaccination schedule and information on the expected lag between the first dose of a vaccine and it becoming effective, the rate that individuals flow from the S compartment to the SV compartment, for example, is
\begin{equation}
    \nu = \frac{\epsilon \, v_d(t)}{N-V(t)}, \label{nu}
\end{equation}
where $v_d(t)$ is the number of vaccinations administered that could become effective at time $t$ (i.e., the number of doses administered $l$ days earlier, where $l$ is the vaccine's lag to effectiveness) and $V(t)$ is the cumulative number of vaccines administered to date, so that $N-V(t)$ is the number of unvaccinated members of the population, from whom each new recipient of a vaccine is assumed to be chosen at random.

The complete set of differential equations to describe this vaccination model has Eqs.~(\ref{1}), (\ref{2}), (\ref{4}) and (\ref{9}) modified as follows to include the flows of effectively vaccinated people:
\begin{align}
\frac{d S}{d t} & = -\lambda S -\nu S\label{1V}\\
\frac{d E}{d t} & =\lambda S - \frac{1}{L}  E-\nu E\label{2V} \\
\frac{d I_a}{d t} & = \frac{f}{L}  E - \frac{1}{D} I_a - \nu I_a\label{4V} \\
\frac{d R}{d t} & = \frac{1}{D} I_a + \frac{1}{D-C+L} I_q+\frac{1}{D-C+L-T} I_{t_2}+\frac{1}{D-C+L} I_{n}-\nu R, \label{9V}
\end{align}
while Eqs.~(\ref{3}), (\ref{5}), (\ref{6}), (\ref{7}) and (\ref{8}) are unchanged (we assume that symptomatic infected individuals are not eligible for vaccination). Note that the force of infection $\lambda(t)$ is given by the modified version in Eq.~(\ref{forceofinfection2}). In addition, the following equations describe the flows in and out of the effectively-vaccinated compartments:
\begin{align}
\frac{d (SV)}{d t} & = \nu S -\lambda\, SV \label{SV}\\
\frac{d (EV)}{d t} & =\nu E +\lambda\, SV - \frac{1}{L}  EV\label{EV} \\
\frac{d (IV)}{d t} & = \nu I_a +\frac{f_2}{L}  EV - \frac{1}{D} IV \label{IV} \\
\frac{d (RV)}{d t} & = \nu R+ \frac{1-f_2}{L} EV + \frac{1}{D} IV. \label{RV}
\end{align}

At the time of writing, there is only very limited clinical evidence regarding the fraction $f_2$ of vaccinated-exposed individuals who can transmit the virus; accordingly, we assume a wide range for this parameter, using a uniform distribution on the interval [0,1], with mean value of 0.5. The multiplicative factor $h_2$ for effective transmission from infected-vaccinated is assumed to be less than the corresponding factor $h$ for transmission from asymptomatic-infected, so the parameter range for $h_2$ is chosen to be uniform on $[0,h]$.

The complete vaccination model above is described by its graph representation in Fig.~\ref{figschematic2} and the results of calibrating and scenario prediction is shown in Fig.~\ref{figvacc}. For clarity we use only a single set of parameters but we illustrate the impact of the vaccination rollout speed by comparing results for scenarios where we assume either $v_d=5000$ or $v_d=10000$ administered vaccines per day. The calibration and scenario match that of Sec.~\ref{sec:scenarios} with an effective contact rate equivalent to $R=0.5$ switching to $R=1.4$ on December 2nd (day 278). The first vaccinations are assumed to become effective on the calibration day 11th November 2020 (day 257). 

\subsection{Reduced-order vaccination model}\label{SMf}
The reduced-order vaccination model has the same compartment structure as the original (no-vaccine) model of   Eqs.~(\ref{1})--(\ref{9}) and Fig.~\ref{fignetwork} but we assume that the effectively-vaccinated fraction $v(t)=\epsilon V(t)/N$ of the population can be applied equally to each of the S, E, etc. compartments to estimate the number of effectively vaccinated individuals of that type (where, as in Eq.~(\ref{nu}), $V(t)$ is the cumulative number of effective vaccinations at time $t$).

Thus, we assume in the reduced model that of the individuals in the S compartment, for example, a fraction $v(t)$ are effectively vaccinated, with similar assumptions on the E, $\text{I}_\text{a}$ and R compartments. We replace the constant parameter $f$ of the original model with a weighted (and time-dependent) version $\tilde f$. The reduced model flows all vaccinated-exposed individuals into the asymptomatic-infected compartment in addition to the fraction $f$ of unvaccinated individuals who also become asymptomatic infected, hence
\begin{equation}
    \tilde f = v+(1-v) f.
\end{equation}
In addition, the reduced model uses a weighted average transmissibility factor $\tilde h$ for the asymptomatic-infected compartment to replace the constant parameter $h$ of the original
model:
\begin{equation}
    \tilde h = \frac{(1-v) f h + v f_2 h_2}{\tilde f}.
\end{equation}
This parameter $\tilde h$ is composed of the weighted transmissibility for the unvaccinated people with asymptomatic infection (contributing $(1-v)f h$ to the numerator) and that of the vaccinated with asymptomatic infection (contributing $v f_2 h_2$ to the numerator).

Replacing $f$ by $\tilde f$ and $h$ by $\tilde h$ in Eqs.~(\ref{1})--(\ref{9}), we run the reduced model and compare with the full vaccination model in Fig.~\ref{figvacc}; we see that the reduced approximation is in very good agreement with the full model. Using the reduced version of the model facilitates the extension of the base model to include, for example, compartments for vaccinations in multiple age-cohorts \cite{Andrade2020}.

\end{document}